\documentclass[doublecol,figures]{epl2} % for 2 columns style with line numbers

\usepackage{amsmath,amssymb}
\usepackage{amsmath,bm}
\usepackage[utf8]{inputenc}

\title{Rumor propagation meets skepticism: a parallel with zombies}
\author{Marco Antonio Amaral\inst{1,2}\thanks{E-mail: \email{marcoantonio.amaral@gmail.com}} 
\and Jeferson J. Arenzon\inst{1}\thanks{E-mail: \email{arenzon@if.ufrgs.br}}}
\institute{
\inst{1}
Instituto de Física, Universidade Federal do Rio Grande do Sul, CP 15051, 91501-970 - Porto Alegre - RS, Brazil\\
\inst{2}
Universidade Federal do Sul da Bahia, 45988 - Praça Joana Angélica 58 - São José, Teixeira de Freitas - BA, Brazil
}
\shortauthor{Amaral \etal}
\date{\today}


\abstract{
We propose a model of rumor spreading in which susceptible, but skeptically oriented individuals may oppose the rumor. Resistance may be implemented either by skeptical activists trying to convince spreaders to stop their activity, becoming stiflers or, passively (non-reactive) as a consequence, for example, of fact-checking. Interestingly, these two mechanisms, when combined, are similar to the (assumed) spreading of a fictitious zombie outbreak, where survivors actively target infected people. We analyse the well-mixed (mean-field) description and obtain the conditions for rumors (zombies) to spread through the whole population. The results show that when the skepticism is strong enough, the model predicts the coexistence of two fixed points (such bistability may be related to polarized situations), with the fate of rumors depending on the initial exposure to it. 
}

\begin{document}
\maketitle
\section{Introduction}

Rumors, like contagious diseases, propagate through contact~\cite{CaFoLo09,PaCaMiVe15}. Although the transmission of the latter is usually involuntary and unconscious, the former involves an intentional action from the spreader, often aimed at a specific target. Despite the important differences between the two processes, there are enough similarities that justify the use of the framework developed to study epidemics in order to better comprehend how an incipient information, whether true or not, may propagate throughout a population~\cite{ArRoRoCoMo17}. 
Understanding how efficiently information diffuses and how the process depends on the contact network and other characteristics of the population may help to somehow predict and control their consequences on the society. Irrespective of its truthfulness, the propagation and subsequent exposure to the information alone might be enough to prime a large number of individuals and shape their future responses~\cite{QuCoLo11,BeCoDaScCaQu15,ViBeZoPeScCaStQu16}. Many examples come from conspiracy theorists (e.g., flat-Earthers), openly anti-science movements (e.g., anti-vaxxers) and, more importantly, political manipulations (by, for example, fake-news crafters and their correlation with the results of elections and referenda~\cite{BeCoDaScCaQu15,ViBeZoPeScCaStQu16,ZoBeViScCaShHaQu17}). Contrary to diseases and vaccines, opposing rumors may be done by skeptically skilled individuals that are able, for example, to check original sources, peer through the available literature and accumulated evidences and neutralize the carrier of the rumors. Rumors may eventually become the prevalent position, disappear or even evolve to a state in which believers and skeptics coexist, in a polarized condition. We explore these possible scenarios in a simple model including
skepticism among the population what also turns out, interestingly, to resemble a zombie outbreak, as we explain below.

Simple models have been important to understand the underlying mechanisms in the above contact processes~\cite{MaDi99}. The three basic dynamical states (compartments) in the first proposed models of rumor propagation (ignorants, spreaders and stiflers)~\cite{GoNe64,DaKe64,DaKe65,GoNe67,MaTh73} map onto those for diseases (susceptible, infected, removed)~\cite{KeMc27}. Further states have been introduced in subsequent generalizations of those models~\cite{CaFoLo09,PaCaMiVe15,Brauer17}. Agents that are unaware of the rumor being propagated, or still uncontaminated by a hazardous strain, are, respectively, ignorant or susceptible (S) to it. 
If the agent receiving the information immediately manifests her skepticism, the attempted transmission may result in the removal of the spreader instead of infecting the susceptible. Otherwise, there is an intermediate, non contagious period of latency after having contact with the rumor, when the exposed agent hesitates and access its plausibility (the E state~\cite{ArSc84,XiJiSoSo15}). 
In this first stage of contamination, 
the agent, once being exposed, is still processing the new information, unsure of its validity, and temporarily refrains from spreading it (equivalently, for a disease, no symptom is observable during the incubation period). 
If the rumor is not convincing, the propagation may halt and the agent becomes a stifler (or removed, R).
Otherwise, it may turn into a rumor spreader (Z) or, in the epidemics context, become infected and develop the contagious symptoms of the disease. 

Many analogies have been drawn between the processes of rumor and disease propagation through contact. In standard disease spreading, the individual being infected has a passive role, all resistance being futile. Although preventive measures may be taken~\cite{VeWiBe16}, they are not directed to a specific spreader and usually have no effect on them. The first skeptical response considered here is performed by the susceptible agent while interacting with the spreader. Despite not having a direct
counterpart in the usual disease contexts, 
such reactive mechanism  is found in recent models for a zombie outbreak~\cite{MuHuImSm09,AlBiMySe15}.
Although belonging to the realm of pop-culture and fantasy, several of its characteristics are similar not only to real epidemics~\cite{CDC} but, as we propose here, to rumor propagation. Following the canon, the zombie condition can be transmitted to healthy humans that, in turn, may also get zombies removed. After the attack, the agent may either die or, once the incubation period is over, become a zombie as well. In close analogy to that, in the rumor propagation model considered here, we study the effects of an immediate-response, skeptically-driven reaction directed toward the spreader.  If not successful, already exposed agents may avoid the spreader condition after critically evaluating the
probability of the rumor to be true, discarding it if not convinced (death in the
zombie scenario) or, otherwise, becoming a spreader.

\section{The model}
\label{section.Model}

\begin{figure}[htb]
\onefigure[width=7cm]{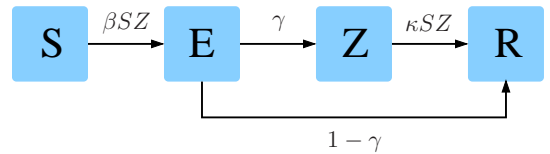}
\caption{Compartments and the corresponding transition rates. When S and Z interact, either S becomes exposed (E) with rate $\beta$ or the Z is removed (with rate $\kappa$), the latter process being driven by the active skepticism of S. Once exposed, the agent may convince herself (with rate $\gamma$) or discard the information ($1-\gamma$) as a consequence, again, of her (passive) skepticism upon reflection.}
\label{fig.SEZR}
\end{figure}

We introduce skepticism at different stages of the infection process when skeptics have the opportunity to critically judge their own opinions after accessing, whenever possible, the 
original sources and, as a consequence, dismiss a rumour when the available evidences are not convincing. When it occurs during the interaction with the spreader, the latter may be removed by the action of a skeptical activist, those that tend to get involved in debunking those propositions lacking or contradicting evidence by directly targeting the sources of misinformation. The second chance occurs when, once exposed, the rumor is spontaneously dismissed (not a reaction to an interaction). We study both mechanisms 
and describe the time evolution of each component of the (infinite) population, assuming homogeneous, well-mixed interactions. The dynamical, mean field equations are: 
\begin{align}
\begin{aligned}
\dot{S}&=-\beta SZ \\  
\dot{E}&=\beta SZ - E \\
\dot{Z}&=\gamma E-\kappa SZ \\  
\dot{R}&=\kappa SZ +(1-\gamma) E.
\label{eq.sezr}
\end{aligned}
\end{align}
A compartment diagram of these interactions is shown in Fig.~\ref{fig.SEZR}. 
The rumor (or the infectious strain) propagates from a knowledgeable spreader/infected Z to an ignorant/susceptible S with rate $\beta$, related to how easy the
rumor propagates, with S becoming exposed (E), albeit not yet contagious.
A fraction $\gamma$ of those exposed eventually becomes spreaders, while
the remaining $1-\gamma$ had enough time to doubt the rumor, becoming removed/stifler (R). 
This passive skepticism is a non reactive, spontaneous mechanism  (i.e., independent of contact with other agents).  Easily spreading rumors  
become ``viral'' while if the information is not interesting or can be easily discarded, upon fact-checking for example, the propagation fades away.  
If transmitted, $Z$ increases, otherwise the spreader is removed, increasing $R$. 
Skepticism is an important, albeit often neglected, ingredient allowing the susceptible population to eliminate the rumor. Its dynamics is different from the classic SIR model (because of the catalytic nature of the removal process)
On the other hand, the removal of a spreader/infected by an S is also possible.
The parameter $\kappa$ is a measure of how intense the skeptical activism is and its capability of counteracting the propagation of misinformation by convincing the spreader.  
 This action depends on the contact between an S and a Z: spreaders/zombies do not become $R$ spontaneously in this model. In this way, the rumor is not automatically accepted and depending on the amount of active skepticism in the population (the parameter $\kappa$), the spreader is persuaded and, consequently, removed. 
This may be also viewed as a very particular antidote mechanism, one in which susceptible agents are responsible for administering a drug and removing infected ones (while, in general, vaccines are devised against the strain, not its carrier).
We assume that the rumor propagation occurs on a fast timescale (compared to the average lifespan of individuals), such that the population demography can be ignored and its size kept fixed.

In the context of epidemiology, if $\gamma$ is very small, most susceptibles that were infected will die before becoming a spreader, which represents the interesting case of a very aggressive strain. By  directly killing the victims, Zs are not replaced, leading to their extinction.  
In this scenario, highly virulent diseases exterminate the whole population before it gets the opportunity to spread out. Or, analogously, when every attempt to spread a very unlikely or unpopular rumor leads to the susceptible person to dismiss it, becoming a stifler.

\section{Results}
\label{section.Results}

\begin{figure}[htb]
\includegraphics[width=0.8\columnwidth]{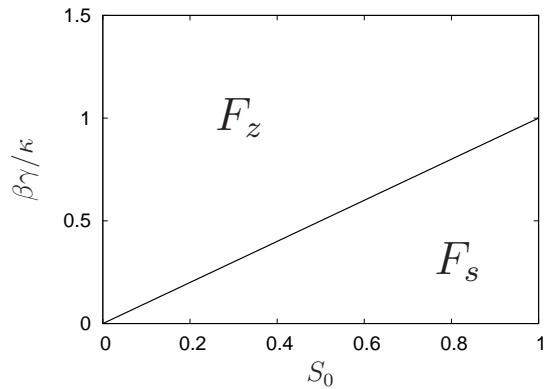}
\caption{For $\beta\gamma>\kappa$,  only the fixed point $F_z$ exists, irrespective of the value 
of $S_0$. For $\beta\gamma<\kappa$, instead, both fixed points are possible but
depend on the initial fraction of susceptibles. The line $S_0=\beta\gamma/\kappa$, on top of which both $S$ and $Z$ decrease as $t^{-1}$,
delimitating the basins of attraction associated with each fixed point.}
\label{fig.diag}
\end{figure}

Besides the normalization condition $S+E+Z+R=1$, there is another integral of motion, $P\equiv (\beta\gamma -\kappa)S+\beta\gamma E+\beta Z$ such that $\dot{P}=0$. These constraints reduce the number of independent variables to two. Eq.~(\ref{eq.sezr}) has two fixed points, both with $E^*=0$ and either
$Z^*=0$ or $S^*=0$: $F_s\equiv (S^*,E^*,Z^*,R^*)=(S^*,0,0,1-S^*)$ and $F_z\equiv (0,0,Z^*,1-Z^*)$. Both $S^*$ and $Z^*$ depend on the initial conditions. For example, considering only susceptible and exposed individuals at $t=0$, $S_0+E_0=1$, we get
\begin{align}
S^* &= \frac{\beta\gamma-\kappa S_0}{\beta\gamma-\kappa} \\
Z^* &= \frac{\beta\gamma-\kappa S_0}{\beta},
\end{align}
for $F_s$ and $F_z$, respectively.
 In the {\it SEZR} simplex, two edges are thus absorbing states: $\overline{RS}$, a mixture of removed and susceptible individuals, and $\overline{ZR}$, with removed and infected/spreaders. In the former, the rumor has faded, and while a fraction $S^*$ of the population never had contact with or remained unconvinced by it, the others became stiflers, $R^*=1-S^*$. In the epidemic context, this means that the infection has been eliminated.
The second fixed point, on the edge $\overline{ZR}$, corresponds to the state where everyone got the infection or got convinced by the rumor, with $R^*$ stiflers and $Z^*$ spreaders. It is not only in the
epidemiological context that this asymptotic state represents an
inconvenient scenario. Indeed, the
rumor had become the norm, a dogma that cannot be debunked due to the absence of skeptical thinkers
among the susceptibles. 
Imposing that both $S_0$ and $S^*$ are in the interval $[0,1]$, we obtain constraints on $F_s$ and $F_z$. For $\beta\gamma>\kappa$, only the fixed point $F_z$ appears, with $Z^*$ in the interval $\gamma-\kappa/\beta\leq Z^*\leq \gamma$, depending on $S_0$. On the other hand, for $\beta\gamma<\kappa$, both fixed points are possible and are related to different intervals of $S_0$ (see Fig.~\ref{fig.diag}). For initial states with $S_0>\beta\gamma/\kappa$, only $F_s$ appears, while the other fixed point, $F_z$, occurs otherwise, i.e., for $S_0<\beta\gamma/\kappa$. Notice, however, that these constraints are not dictated by stability considerations. 

Analysing the Jacobian of Eq.~(\ref{eq.sezr}), the stability of the fixed points can be studied.
For $F_s$, the eigenvalues are either 0 or
\begin{equation}
 \lambda_s^{\pm}= -\frac{\kappa S^*+1}{2} \pm\frac{1}{2} \sqrt{(\kappa S^*-1)^2+4S^*\gamma \beta}.
\end{equation} 
While $\lambda_s^-$ is always negative, $\lambda_s^+$ is positive if $ \beta\gamma > \kappa$ (a region where $S^*$ is not in the interval $[0,1]$ anyway), and negative otherwise. On the other hand, for $F_z$, the eigenvalues are either null or negative. 
Both fixed points are non hyperbolic (because of the zero eigenvalues) and 
the linear stability analysis is not enough to elucidate the actual nature of each region in Fig.~\ref{fig.diag}.
Nonetheless, the numerical integration of Eq.~(\ref{eq.sezr}) may get a hint on whether the orbits are attracted or not to these fixed points.

\begin{figure}[htb]

\includegraphics[width=0.325\columnwidth]{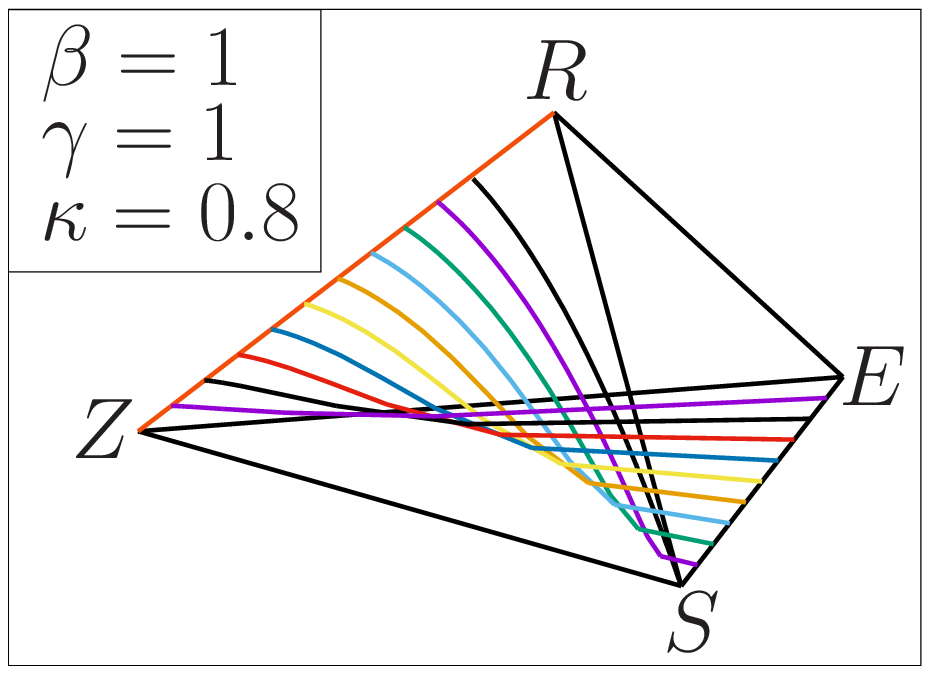}
\includegraphics[width=0.325\columnwidth]{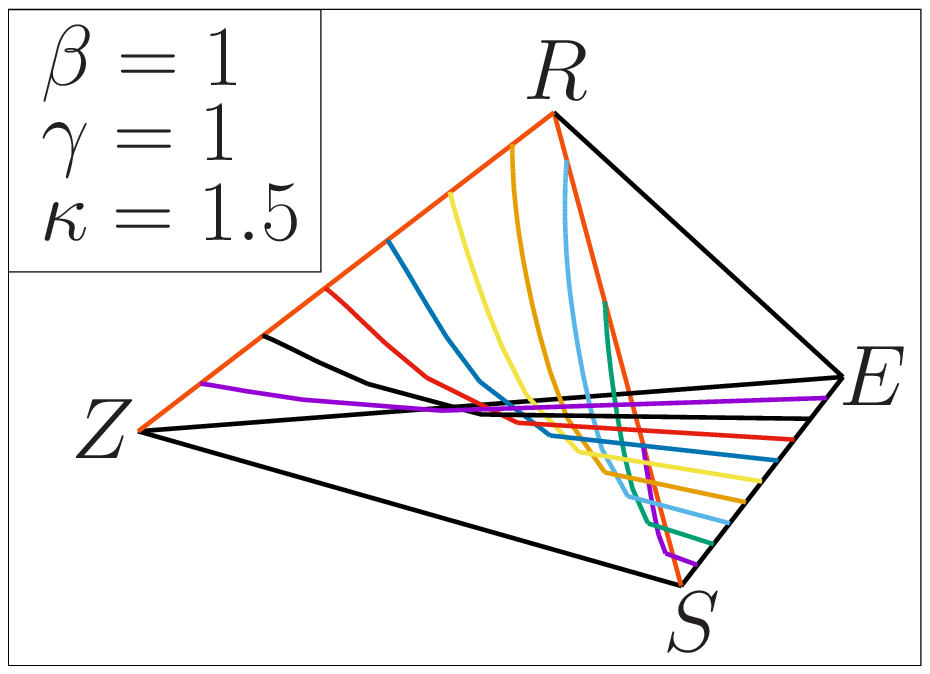}
\includegraphics[width=0.325\columnwidth]{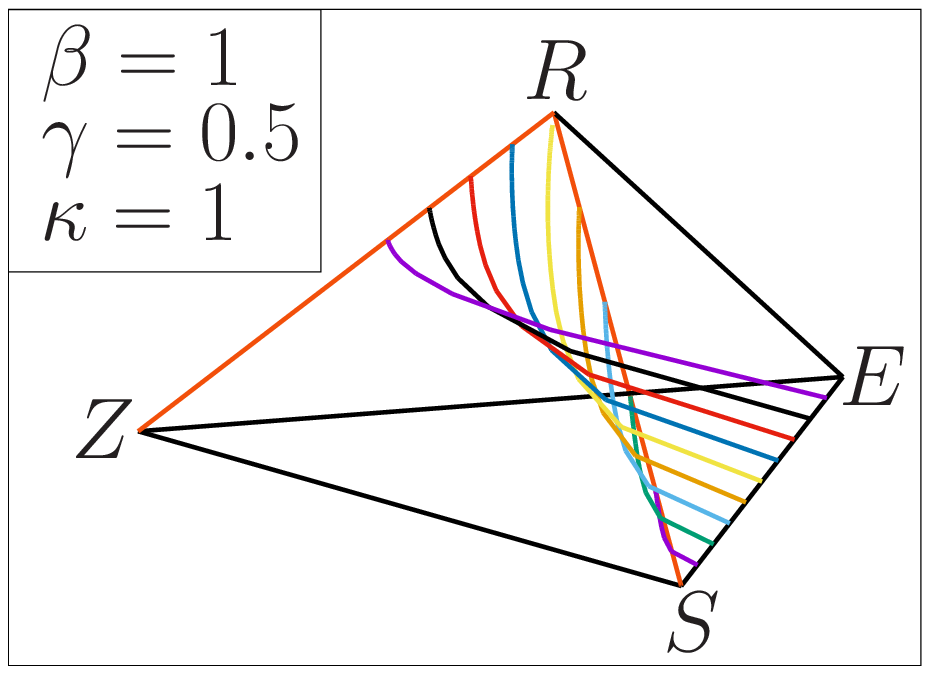}

(a) \hspace{2.2cm} (b) \hspace{2.2cm} (c) 

\caption{Simplex trajectories for several values of the parameters, for both
$\beta\gamma>\kappa$ (a) and $\beta\gamma<\kappa$ (b and c).
In all cases the initial conditions are evenly distributed on $\overline{SE}$. For $\beta\gamma>\kappa$  (a), all asymptotic states are on the $\overline{ZR}$ edge (orange).
As $\kappa$ increases, one finds bistability as part of the initial conditions lead to asymptotic states on the $\overline{RS}$ edge (orange) as well (b). When $\kappa$ becomes larger than $\beta\gamma$, both
fixed points coexist (bistability) and the asymptotic state depends on $S_0$.
By increasing $\gamma$, on the other hand, the allowed interval on
$\overline{EZ}$ decreases because $Z^*\leq \gamma$.} 
\label{fig.simplex}
\end{figure}

Fig.~\ref{fig.simplex} shows several trajectories, obtained by numerically  
integrating  Eq.~(\ref{eq.sezr}), starting with different mixtures of exposed and susceptible individuals ($S_0+E_0=1$, i.e., on the edge $\overline{SE}$ of the 3-simplex).  If $\beta\gamma >\kappa$ (Fig.~\ref{fig.simplex}a), all orbits are attracted to $F_z$, a mixed state fixed point on the $\overline{ZR}$ edge with $\gamma-\kappa/\beta\leq Z^*\leq\gamma$ infected and $1-Z^*$ removed individuals. 
Despite the existence of zero eigenvalues, the numerical integration indicates that $F_z$ 
is indeed asymptotically stable for $\beta\gamma >\kappa$. 
The larger the number of initially exposed agents is, the smaller the number of removed ones in the asymptotic state (because removal is done by susceptibles). 
The regime for $\beta\gamma <\kappa$ is richer, the system presenting bistability:
depending on the initial state, the system may end up either on the
$\overline{RS}$ ($F_s$) or the $\overline{ZR}$ ($F_z$) edge. 
Indeed, for $S_0<\beta\gamma/\kappa$,  the orbit
is attracted to $F_z$ while if $S_0>\beta\gamma/\kappa$, the asymptotic
state is on $F_s$ (see Fig.~\ref{fig.diag}). For $S_0=\beta\gamma/\kappa$, $S^*=Z^*=0$ and everyone in
the population is eventually removed (we observe, numerically, that the approach to the asymptotic state, in this case, is power-law,
$t^{-1}$). 
A smaller (larger) initial fraction of exposed agents increases the number of infected (susceptible) individuals in the asymptotic state, while an intermediate $E_0$ gives a maximum of removed ones.   Notice that except for the trivial
$\gamma=0$ case where $S^*=S_0$, nowhere in the space of parameters all initial conditions lead
to $F_s$. 
This result differs from the original SZR model, that does not present bistability: susceptible individuals always survive if $\kappa>\beta$, otherwise zombies always succeed~\cite{MuHuImSm09,AlBiMySe15,WiBl13}. Since $Z^*=\gamma-\kappa S_0/\beta\leq \gamma$, its maximum
value occurs for $S_0\to 0$ (see Fig.~\ref{fig.simplex}c): every agent is already exposed and a fraction $\gamma$ becomes infected. 
Clearly, even in a favorable situation for the spreaders, a low value of $\gamma$ will be detrimental to them.

A complementary approach to the stability of the fixed points is through the basic reproductive number ${\cal R}_0$,  the number of secondary infections that a single infected individual can generate before it dies or becomes cured (see, e.g., Ref.~\cite{BrCa12,Brauer17} and references therein). Thus, starting from a disease-free equilibrium state, if ${\cal R}_0>1$, there is an endemic state with a finite fraction of  infected agents, whereas if ${\cal R}_0<1$, the infection eventually disappears. 
We first define $F_i=\{\beta S Z, 0\}$, the  rate that secondary infections increase the $i$-th disease compartment, and $V_i=\{E,\kappa S Z- \gamma E\}$, the rate that disease progression, death and recovery decrease it. The next generation matrix is ${\bm K}= \bm{FV }^{-1}$, where
$F_{ij}=(\partial F_i/\partial x_j)_s$ and $V_{ij}=(\partial V_i/\partial x_j)_s$ are evaluated at $F_s$, the disease-free equilibrium state. Thus:
\begin{equation}
 {\bm K}=
\begin{bmatrix}
   0 & \beta S^* \\
   0 & 0   
\end{bmatrix}
\begin{bmatrix}
  1 & 0  \\
 -  \gamma  & \kappa S^* 
\end{bmatrix}^{-1}
=
\begin{bmatrix}
  \beta \gamma / \kappa & \beta / \kappa  \\
 0  &  0
\end{bmatrix}
\end{equation} 
where the element $ij$ gives the expected number of secondary infections in the compartment $i$ produced by individuals initially in compartment $j$. The eigenvalues are
$\lambda=0$ and $\beta\gamma / \kappa$ and 
${\cal R}_0=\beta \gamma / \kappa$. As a consequence, in agreement with the previous analysis, if $\beta \gamma>\kappa$ there is an epidemic state, otherwise the infection disappears.

\begin{figure}[hbt]
\includegraphics[width=8cm]{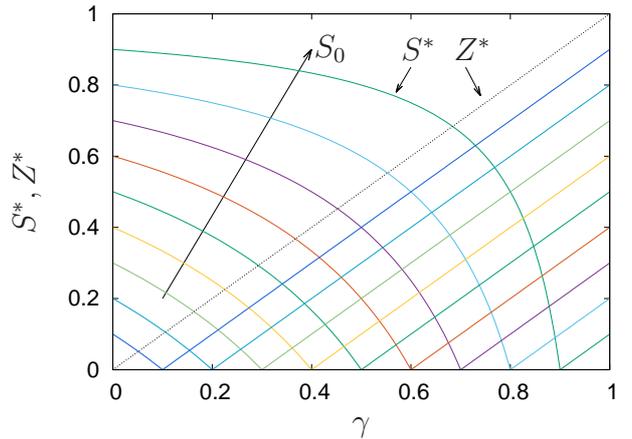}
\caption{Asymptotic fraction of susceptible ($S^*$, curved lines) and spreaders ($Z^*$, straight lines), as a function of $\gamma$, for different initial values of $S_0=1-E_0$ and $\kappa=\beta$.}
\label{fig.asympt_g}
\end{figure}

Consider first the case $\gamma=1$, where passive skepticism is absent (all individuals, once exposed, eventually become spreaders). When spreaders are more efficient than skeptics ($\beta>\kappa$), only the $F_z$ fixed point is possible and, as expected, asymptotically there are more spreaders, $Z^*=1-(\kappa/\beta)S_0$, than initially exposed agents, $E_0=1-S_0$. Under these conditions, the larger $E_0$ is, the greater the asymptotic number of spreaders. Thus, against efficient spreaders, the best action is, obviously, to reduce the initial exposition to the rumor. However, the presence of passive, non-reactive skeptics ($\gamma<1$), not only reduces the propagation of the rumor, $Z^*=\gamma-(\kappa/\beta)S_0$, but defines a threshold for $S_0$: for $S_0<\beta\gamma/\kappa$, the asymptotic state is $F_z$, otherwise it is $F_s$ (too many initially exposed agents lead to infecting all susceptibles). Another interesting case is when spreaders and skeptics are 
equally efficient ($\beta=\kappa$) in changing each other strategy (to E and R, respectively) but, by changing $\gamma$, the probability of the rumor be considered true varies
(as $\gamma$ approaches zero, the rumor becomes highly improbable). 
Fig.~\ref{fig.asympt_g} shows that, the larger the initial exposition $E_0$ is,
the smaller is the threshold value of $\gamma$ above which $Z^*$ is the asymptotic
state. Thus, despite the effort of skeptical activists, stopping the propagation
of a rumor may eventually come from the passive skepticals
(in the zombie analogy, a too aggressive atack that kills more than infect people
will decrease the final number of infected).

\section{Conclusions}
\label{section.Conclusions}

A fundamental question regarding rumor propagation is how to deal with conspiracies, fake-news and other persistent ideas against which evidences abound. Here, at variance with the usual diseases, the spreader may be actively opposed by the same very agents that are susceptible to the rumor. 
By behaving as a skeptical activist, an ignorant/susceptible S that interacts with a spreader Z and does not get convinced, may counter-argument and dissuade the latter, that thus refrains from further spreading. Otherwise, S becomes exposed to the rumor, albeit delaying its decision on whether believe in it or not. Both cases, the suppression (Z $\to$ R) and the initial acceptance (S $\to$ E) of the rumor depend, in our model, on the interaction (reactive terms) between susceptibles and spreaders. The exposed agent eventually decides, spontaneously, between accept and propagate the rumor (E $\to$ Z) or stop it (E $\to$ R), stepping out from the process (this may also occur because of apathy~\cite{BoMeGoMo13}). In the latter transition, the skepticism  is passive, a consequence of self-convincing upon reflection and fact-checking, and does not involve  any interactions whatsoever. Therefore, the main question is on what conditions it is more convenient to directly attack a rumor and its messenger (what may backfire, increasing its public exposition and the attention it gets) or quietly dismiss it. Or, in the framework of our model, is it better to increase $\kappa$ or decrease $\gamma$, and when?

For some range of the parameters, the model presents bistability, where both fixed points are attractive within the corresponding basin of attraction. This suggests that when spatial correlations become important, beyond the well-mixed, mean-field approximation, heterogeneities in the initial state may persist, leading to the competition between both fixed points. Domains associated with each of them may evolve in time, similarly to coarsening physical systems, and both strategies may coexist for long periods of time, a process that mimics the polarization and competition between bubbles of consensus. Despite such polarization, the persistent coexistence between S and Z may be interpreted as a tolerant scenario where different opinions are present in the population. %What are the mechanisms allowing such state is an important question.

We can obtain the conditions for the rumor to either reach the whole population or disappear. Specifically, we find that, the level of skepticism of a population can limit the final fraction of rumor spreaders even when there are no susceptible individuals left. At the same time we find that the rate at which these susceptible individuals fight rumor spreaders is a very important parameter. Based on it, the asymptotic behavior of the population can change from a coexistence state of spreaders and removed individuals to one with susceptible and removed individuals.

Further generalizations are possible and may lead to improved strategies to face threats from the dissemination of unproved claims.  For example, agents recovering after being removed, or the inclusion of demographic effects, may help avoiding the absorbing states, keep some degree of coexistence in the now fluctuating population and perhaps unveil new phenomena. Indeed, by reinserting stiflers back into the susceptible compartment~\cite{HoWa14} one obtains a cyclic model with four possible strategies (see Refs.~\cite{LuRiAr13,RuAr14} and references therein) where other coexisting states may be possible. Once dissuaded, the spreader may also become a zealot, an individual whose role is to combat the rumor, as the skeptical activists, but without the risk of becoming infected, thus differing from susceptibles. With the price of including one further parameter in the model, it is possible to control the timescale of the exposed period by replacing $-E$ by $-\sigma E$ in the second line of Eq.~(\ref{eq.sezr}).
This may be useful and interesting for the dynamics of the rumor propagation but does not alter the asymptotic states and the  conditions to attain them obtained above.
Notice that it is possible to reinterpretate the model, replacing skepticals by deniers (or fake-news propagators) trying to replace valid, well established knowledge and legitimate fringe positions (the war on evolution is an example). It would be interesting to model a population composed by these two possible positions regarding the validity of the rumor, having the verisimilitude as a possible ingredient breaking the symmetry between them. Furthermore, the assumption of homogeneous mixing is a strong simplification. Heterogeneities may
arise from largely different densities (for example, at the beginning or end of an epidemy), disordered, non homogeneous landscapes or contact networks. In particular, it would be important to consider the spread phenomena on correlated, heterogeneous structures more akin to the complex, real-world networks~\cite{ReBi07,FeCaPa12,PaCaMiVe15,SuJuJiWaWa16}. Heterogeneities are also known to occur inside the echo chambers of a polarized condition~\cite{BeCoDaScCaQu15}, signalling that the simple idea of homogeneous compartments widely used in epidemiological models is an initial approximation. Another implicit simplification in our model is to consider only simple propagation (the activation of a node is triggered by a single active neighbor) while more ellaborate information in the real world involves cooperative, complex propagation~\cite{CeMa07,CeEgMa07} where multiple active neighbors are necessary to allow further diffusion.
These are examples of open problems to be considered in the future.

Interestingly, the mechanism to direct and individually counteract the propagation of rumors by skeptical agents is similar to the one depicted in popular scenarios for a zombie outbreak. Such descriptions, whose relevance relies on the possibility of tracing parallels and analogies, have a strong popular appeal~\cite{CDC}. 
While most epidemic models suppose that infected individuals spontaneously becomes removed, rumors often find resistance. Thus, in order to understand how skeptical a population should be to stop rumor spreading, we may get insight from the active confrontation between survivors and zombies. This is an important and urgent issue as the zombie apocalypse, albeit disguised as pseudosciences and fake-news spreading, is already upon us.  

\acknowledgments
MAA was partially supported by the Brazilian agency CNPq, project 437983/2016. JJA
acknowledges Antonio Scala for comments on the manuscript and the partial support from the INCT-Sistemas Complexos, CNPq (projects 423283/2016 and 308927/2017) and CAPES (Finance Code 001).

%\bibliographystyle{eplbib} 
%\bibliography{bio.bib} 

\begin{thebibliography}{10}
\expandafter\ifx\csname url\endcsname\relax\def\url#1{\texttt{#1}}\fi

\bibitem{CaFoLo09}
\Name{Castellano C., Fortunato S. \and Loreto V.} \REVIEW{Rev. Mod.
  Phys.}{81}{2009}{591}.

\bibitem{PaCaMiVe15}
\Name{Pastor-Satorras R., Castellano C., Van~Mieghem P. \and Vespignani A.}
  \REVIEW{Rev. Mod. Phys.}{87}{2015}{925}.

\bibitem{ArRoRoCoMo17}
\Name{de~Arruda G.~F., Rodrigues F.~A., Rodríguez P.~M., Cozzo E. \and Moreno
  Y.} \REVIEW{J. Complex Netw.x}{x}{2017}{cnx024}.

\bibitem{QuCoLo11}
\Name{Quattrociocchi W., Conte R. \and Lodi E.} \REVIEW{Adv. Comp.
  Syst.}{14}{2011}{567}.

\bibitem{BeCoDaScCaQu15}
\Name{Bessi A., Coletto M., Davidescu G.~A., Scala A., Caldarelli G. \and
  Quattrociocchi W.} \REVIEW{PLoS ONE}{10}{2015}{e0118093}.

\bibitem{ViBeZoPeScCaStQu16}
\Name{Del~Vicario M., Bessi A., Zollo F., Petroni F., Scala A., Caldarelli G.,
  Stanley H.~E. \and Quattrociocchi W.} \REVIEW{Proc. Nat. Acad.
  Sci.}{113}{2016}{554}.

\bibitem{ZoBeViScCaShHaQu17}
\Name{Zollo F., Bessi A., Del~Vicario M., Scala A., Caldarelli G., Shekhtman
  L., Havlin S. \and Quattrociocchi W.} \REVIEW{{PLoS
  ONE}}{12}{2017}{e0181821}.

\bibitem{MaDi99}
\Name{Marro J. \and Dickman R.} \Book{Nonequilibrium phase transitions in
  lattice models} (Cambridge University Press, Cambridge) 1999.

\bibitem{GoNe64}
\Name{Goffman W. \and Newill V.~A.} \REVIEW{Nature}{204}{1964}{225}.

\bibitem{DaKe64}
\Name{Daley D.~J. \and Kendall D.~G.} \REVIEW{Nature}{204}{1964}{1118}.

\bibitem{DaKe65}
\Name{Daley D.~J. \and Kendall D.~G.} \REVIEW{IMA J. Appl. Math.}{1}{1965}{42}.

\bibitem{GoNe67}
\Name{Goffman W. \and Newill V.~A.} \REVIEW{Proc. R. Soc. A}{298}{1967}{316}.

\bibitem{MaTh73}
\Name{Maki D.~P. \and Thompson M.} \Book{Mathematical models and applications}
  (Prentice-Hall Inc., Englewood Cliffs, N.J.) 1973.

\bibitem{KeMc27}
\Name{Kermack W.~O. \and Mc{K}endrick A.~G.} \REVIEW{Proc. R. Soc.
  A}{115}{1927}{700}.

\bibitem{Brauer17}
\Name{Brauer F.} \REVIEW{Infect. Dis. Mod.}{2}{2017}{113}.

\bibitem{ArSc84}
\Name{Aron J.~L. \and Schwartz I.~B.} \REVIEW{J. Theor. Biol.}{110}{1984}{665}.

\bibitem{XiJiSoSo15}
\Name{Xia L.-L., Jiang G.-P., Song B. \and Song Y.~R.} \REVIEW{Physica
  A}{437}{2015}{295}.

\bibitem{VeWiBe16}
\Name{Verelst F., Willem L. \and Beutels P.} \REVIEW{J. R. Soc.
  Interface}{13}{2016}{20160820}.

\bibitem{MuHuImSm09}
\Name{Munz P., Hudea I., Imad J. \and Smith? R.~J.} \Book{When zombies attack!:
  {M}athematical modelling of an outbreak of zombie infection} in
  \Book{Infectious Disease Modelling Research Progress}, edited by
  \Name{Tchuenche J.~M. \and Chiyaka C.} (Nova Science Publishers, NY USA) 2009
  Ch.~4 pp. 133--150.

\bibitem{AlBiMySe15}
\Name{Alemi A.~A., Bierbaum M., Myers C.~R. \and Sethna J.~P.} \REVIEW{Phys.
  Rev. E}{92}{2015}{052801}.

\bibitem{CDC}
\Name{{U. S. Department of Health, Human Services Centers for Disease Control,
  and Prevention}} \Book{Preparedness 101: Zombie pandemic}
  http://www.cdc.gov/phpr/zombies (2011).
\newline\url{http://www.cdc.gov/phpr/zombies}

\bibitem{WiBl13}
\Name{Witkowski C. \and Blais B.} \Book{Bayesian analysis of epidemics -
  zombies, influenza, and other diseases} arXiv:1311.6376 (2013).

\bibitem{BrCa12}
\Name{Brauer F. \and Castillo-Chavez C.} \Book{{Mathematical Models in
  Population Biology and Epidemiology}} Texts in Applied Mathematics (Springer,
  New York) 2012.

\bibitem{BoMeGoMo13}
\Name{Borge-Holthoefer J., Meloni S., Gonçalves B. \and Moreno Y.} \REVIEW{J.
  Stat. Phys.}{151}{2013}{383}.

\bibitem{HoWa14}
\Name{Hochreiter R. \and Waldhauser C.} \Book{Zombie politics: Evolutionary
  algorithms to counteract the spread of negative opinions} arXiv:1401.6420
  (2014).

\bibitem{LuRiAr13}
\Name{Lütz A.~F., Risau-Gusman S. \and Arenzon J.~J.} \REVIEW{J. Theor.
  Biol.}{317}{2013}{286}.

\bibitem{RuAr14}
\Name{Rulquin C. \and Arenzon J.~J.} \REVIEW{Phys. Rev. E}{89}{2014}{032133}.

\bibitem{ReBi07}
\Name{Real L.~A. \and Biek R.} \REVIEW{J. R. Soc. Interface}{4}{2007}{935}.

\bibitem{FeCaPa12}
\Name{Ferreira S.~C., Castellano C. \and Pastor-Satorras R.} \REVIEW{Phys. Rev.
  E}{86}{2012}{041125}.

\bibitem{SuJuJiWaWa16}
\Name{Sun G.-Q., Jusup M., Jin Z., Wang Y. \and Wang Z.} \REVIEW{Phys. Life
  Rev.}{19}{2016}{43}.

\bibitem{CeMa07}
\Name{Centola D. \and Macy M.} \REVIEW{Am. J. Sociol.}{113}{2007}{702}.

\bibitem{CeEgMa07}
\Name{Centola D., Eguíluz V.~M. \and Macy M.~W.} \REVIEW{Physica
  A}{374}{2007}{449}.

\end{thebibliography}

\end{document}